\documentstyle[aps,epsf,epsfig]{revtex}

\begin{document}
\author{D. Foerster and Y. Meurdesoif, CPTMB, URA1537}
\address{Universit\'{e} de Bordeaux\\
Rue du Solarium, 33174 Gradignan, France}
\author{B. Malet, Ecole Normale Superieure de Lyon}
\address{46 All\'{e}e d'Italie, 69364 Lyon, France}
\title{Magnetic Structure and NMR signal of Spin Peierls Solitons}
\maketitle

\begin{abstract}
We compute the magnetic profile of spin Peierls solitons in a simple
Heisenberg model with magneto elastic couplings, using independently the
DMRG method of White and the Hartree Fock approximation. The results
obtained in such a static model are incompatible with existing NMR data on
CuGeO, but the distribution of averaged spins agrees qualitatively with
the data. We conclude that the dynamics in the
spin plus lattice system must be included 
in a more detailed theory of the spin Peierls transition in CuGeO.
\end{abstract}

\section{Experimental motivation, model and technique of calculation}

Spin Peierls compounds such as CuGeO have solitons in their incommensurate
phase \cite{Boucher} the magnetic profile of which can be measured by NMR
techniques \cite{Berthier} and which must be accounted for by any
theoretical model. Recently a systematic study of solitons in the Heisenberg
spin Peierls model was done by Quantum Monte Carlo methods\cite{Feiguin},
with results that differ essentially from those obtained in the xy model\cite
{exactxy}. However, these authors did not determine the distribution of
spins and the corresponding NMR signal in the incommensurate phase, which
are the subject of the present paper.

Here we calculate the magnetic structure of solitons in the Heisenberg Spin
Peierls model via the DMRG method of White \cite{White} and also within the
Hartree Fock approximation \cite{Fock}. Our (essentially exact) DMRG results
in the dimerised phase agree with those of ref\cite{Feiguin} and they are
also surprisingly close to the results of a Hartree Fock calculation.

The magnetic profiles in the incommensurate phase that we compute by
Hartree Fock disagree with those reconstructed from NMR data \cite{Berthier}.
However, if we eliminate the oscillations in these profiles by averaging over
even and odd sites, we find spin distributions that agree qualitatively with
those obtained by NMR
in\ ref[\cite{Berthier}].

The spin gap Peierls in CuGeO and similar materials \cite{Boucher} may be
due to (i) magnetoelastic couplings \cite{Giesskanne} and/or (ii)
interactions among the chains \cite{Tsvelik}. We consider the simplest
realisation of case (i): 
\begin{eqnarray}
H &=&\sum_{n}J_{n,n+1}\overrightarrow{s}_{n+1}\overrightarrow{s}_{n}+\frac{%
\kappa }{2}\sum \left( \Delta J_{n+1,n}\right) ^{2}+\mu B\sum_{n}s_{n}^{z}
\label{SpinPeierlsModel} \\
J_{n,n+1} &=&1+\Delta J_{n,n+1}  \nonumber
\end{eqnarray}
$\overrightarrow{s}_{n}$ are spin $1/2$ operators, all energies are measured
in units of $J$ and $\kappa $ is the only parameter of this model.
The $\Delta J_{n,n+1}$ are due to elastic deformations $\Delta J_{n,n+1}\sim
(x_{n}-x_{n+1})$ the fluctuations of which are thought to be stabilised by
inter chain interactions and which are treated classically. This approximation 
ignores the
competing energy scales of phonons \cite{Braden} and magnons for CuGeO 
(an attempt to go beyond the static approximation was undertaken in, for example, 
\cite{Uhrig}). We
also ignored second nearest neighbour exchange couplings that might be
needed in a realistic model of CuGeO \cite{Buechner}.
The stationary point of $<H>$ with respect to the deformations $\Delta
J_{n,,n+1}$ is given by 
\begin{eqnarray}
\Delta J_{n,,n+1} &=&-\frac{1}{\kappa }<\overrightarrow{s}_{n}%
\overrightarrow{s}_{n+1}>+const  \label{elastic} \\
\sum \Delta J_{n,n+1} &=&0  \nonumber
\end{eqnarray}
and this inhomogenous classical background $\Delta J_{n,n+1}$ is the key
difficulty in tackling the Spin Peierls Hamiltonian of eq(\ref
{SpinPeierlsModel}). We use the DMRG method of White in the original spin
variables, while for Hartree Fock we use the Jordan-Wigner transform 
\cite{JordanWigner} of eq(\ref{SpinPeierlsModel}): 
\begin{eqnarray}
s_{n}^{+} &=&\prod_{k=1}^{n-1}\left( \psi _{k}^{+}\psi _{k}-\psi _{k}\psi
_{k}^{+}\right) *\psi _{n}^{+}\mbox{, }s_{n}^{z}=\frac{1}{2}\left( \psi
_{n}^{+}\psi _{n}-\psi _{n}\psi _{n}^{+}\right)  \label{JordanWigner} \\
H_{spin} &=&\sum_{n}J_{n+1,n}*\left( \frac{1}{2}\psi _{n+1}^{+}\psi
_{n}+h.c.+s_{n}^{z}s_{n+1}^{z}\right) +\mu B\sum_{n}s_{n}^{z}  \nonumber
\end{eqnarray}
As usual, the Hartree Fock (HF) approximation \cite{Fock} is given by an
energy functional

\begin{eqnarray}
E(\rho ) &=&<H_{spins}>  \label{HartreeFock} \\
&=&\sum J_{n,n+1}\left[ (s_{n+1}^{z}s_{n}^{z}-(\rho _{n,n+1}-\frac{1}{2}%
)(\rho _{n+1,n}-\frac{1}{2})+\frac{1}{4}\right] +\mu B\sum s_{n}^{z}\mbox{ }
\nonumber \\
\mbox{ }s_{n}^{z}\mbox{ } &=&\rho _{n,n}-\frac{1}{2}  \nonumber
\end{eqnarray}
plus a consistency condition 
\begin{equation}
\rho _{mn}=<\psi _{m}^{+}\psi _{n}>\mbox{, }H_{eff}=\sum_{m,n}\left( \frac{%
\delta E}{\delta \rho }\right) _{mn}\psi _{m}^{+}\psi _{n}
\label{selfconsistent}
\end{equation}

\section{First tests of our methods}

To test our DMRG \cite{White} and HF approaches, we consider the ground
state energy of eq(\ref{SpinPeierlsModel}) for prescribed alternating
dimerisation $\Delta J$ at $B=T=\kappa =0$. In the HF calculation there is
an oscillating effective coupling $K_{n,n+1}=K+\Delta K(-)^{n}$ that
reflects oscillations in $J_{n,n+1}=1+\Delta J(-)^{n}$ and that includes
oscillations of the Hartree Fock variables. We find (details are given in an
appendix) 
\begin{eqnarray}
K &=&s+\Delta s\Delta J\text{, }\Delta K=s\Delta J+\Delta s
\label{HFresults} \\
\Delta s &=&\frac{2\Delta K}{\pi K}{\cal D}\text{, }s=1+\frac{2}{\pi }%
\left( {\cal K}-{\cal D}\right)  \nonumber \\
\frac{E(\rho )}{volume} &=&-\frac{1}{4}\left( s^{2}+\Delta s^{2}+2\Delta
Js\Delta s-1\right)  \nonumber
\end{eqnarray}
where ${\cal D}$ ,${\cal K}$ are standard elliptic functions with modulus 
$\kappa =%
\sqrt{1-\left( \frac{\Delta K}{K}\right) ^{2}}$ . Figure(1) displays,
descending in energy, (i) the ground state energy of the renormalised xy
model 
\begin{equation}
H_{xy}=(1+\frac{1}{\pi }%
)\sum_{n}J_{n+1,n}(s_{n}^{x}s_{n+1}^{x}+s_{n}^{y}s_{n+1}^{y})  \label{xy}
\end{equation}

(ii) the Hartree-Fock energy and (iii) the DMRG result (DMRG reproduces the
Bethe Ansatz result $E_{0}=1/4-\log 2$ at $\Delta J=0 $).

\begin{figure}
\mbox{   \epsfig{file=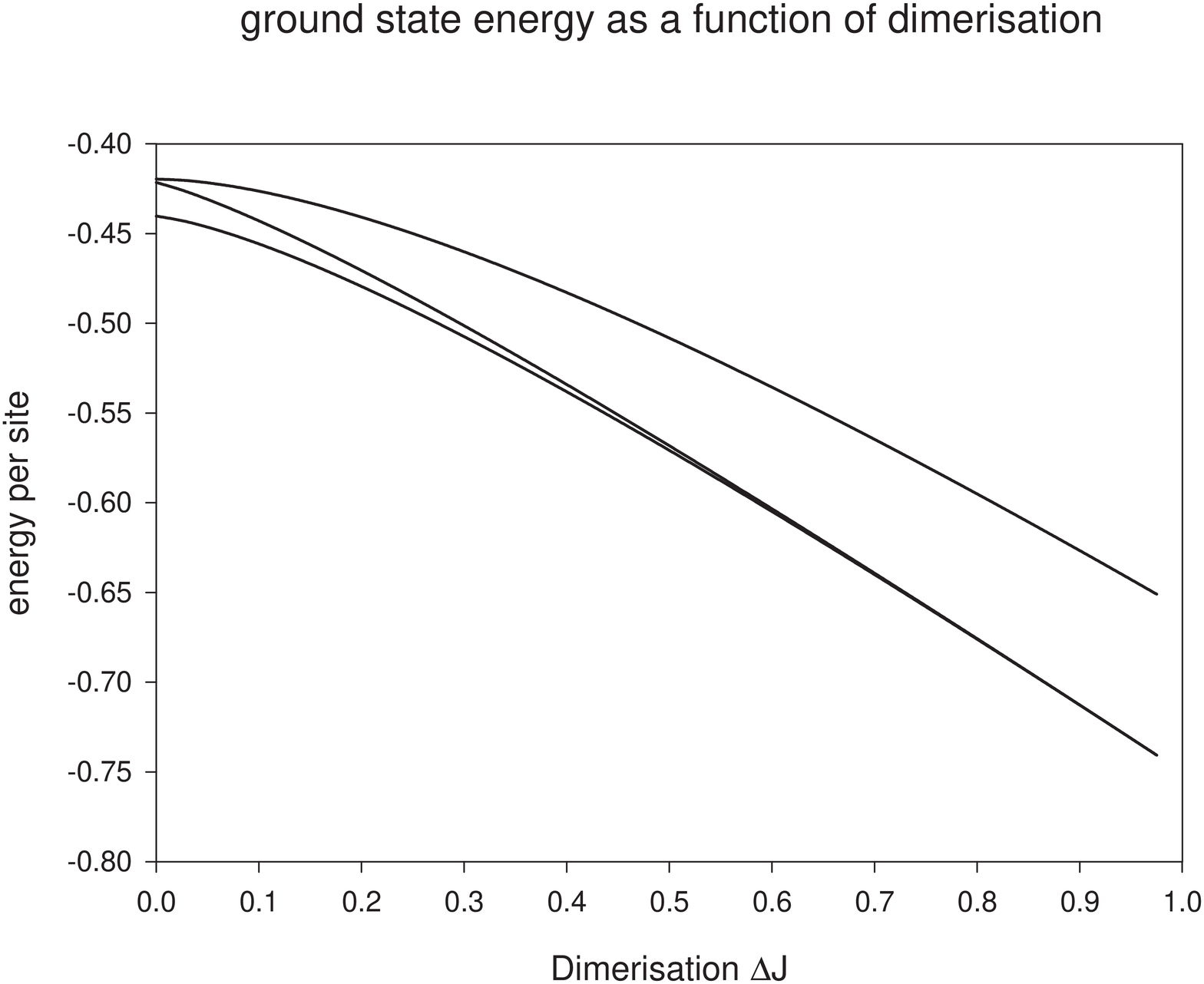,width=10cm}}
\caption{energy per site of a dimerised Heisenberg chain according to, 
         from top to bottom, 
         renormalised xy model, HF approximation and DMRG}
\end{figure}

The DMRG and Hartree Fock results are close and at $\Delta J=1$ where the
system decouples into independent dimers, the Hartree Fock approximation is
exact. Less satisfactory is the incorrect gap of the Hartree-Fock
result at $\Delta J=0$ that was avoided by Bulaevski\cite{Giesskanne}
by assuming the Hartree Fock parameter to be uniform along the chain. We
conclude that the Hartree Fock approximation is distinct from the
renormalised xy model.

Both in DMRG and HF we iterate eq(\ref{elastic}) to converge towards the
fixed point. We test this procedure on the $s=1/2$ solitons of the xy spin
Peierls model of eq(\ref{xy}) or, equivalently the charge=$1/2$ solitons of
the Peierls model of polyacetylene \cite{polyacetylene}. We recall that in
the xy Spin Peierls model $<s_{x}^{z}>=0$ for half the points $[(L+1)/2]$ on
a chain of odd length $L$ with periodic boundary conditions \cite
{LatticeIndex}. This is due to the fact that the Hamiltonian couples only
even with odd points and is of the form

\begin{equation}
H_{xy}=\psi ^{+}M\psi \mbox{, }M=\left( 
\begin{array}{ll}
0 & A \\ 
A^{+} & 0
\end{array}
\right)
\end{equation}
with a spectrum that is symmetric under $E\rightarrow -E:$%
\begin{equation}
\left( 
\begin{array}{l}
\psi _{e} \\ 
\psi _{o}
\end{array}
\right) _{E}\longleftrightarrow \left( 
\begin{array}{l}
\psi _{e} \\ 
-\psi _{o}
\end{array}
\right) _{-E}  \nonumber
\end{equation}
By the symmetry $E\leftrightarrow -E$ there are $2L$ eigenvalues of $\pm E$
plus the eigenvalue $E=0$ of the form 
\begin{equation}
\psi _{E=0}=\left( 
\begin{array}{l}
0 \\ 
\psi _{o}
\end{array}
\right)  \nonumber
\end{equation}
which alone contributes to $<s_{x}^{z}>$ and which is responsible for the
vanishing of $<s_{x}^{z}>$ on half the lattice. All this is borne out by
figure 2:

\begin{figure}
\mbox{\epsfig{file=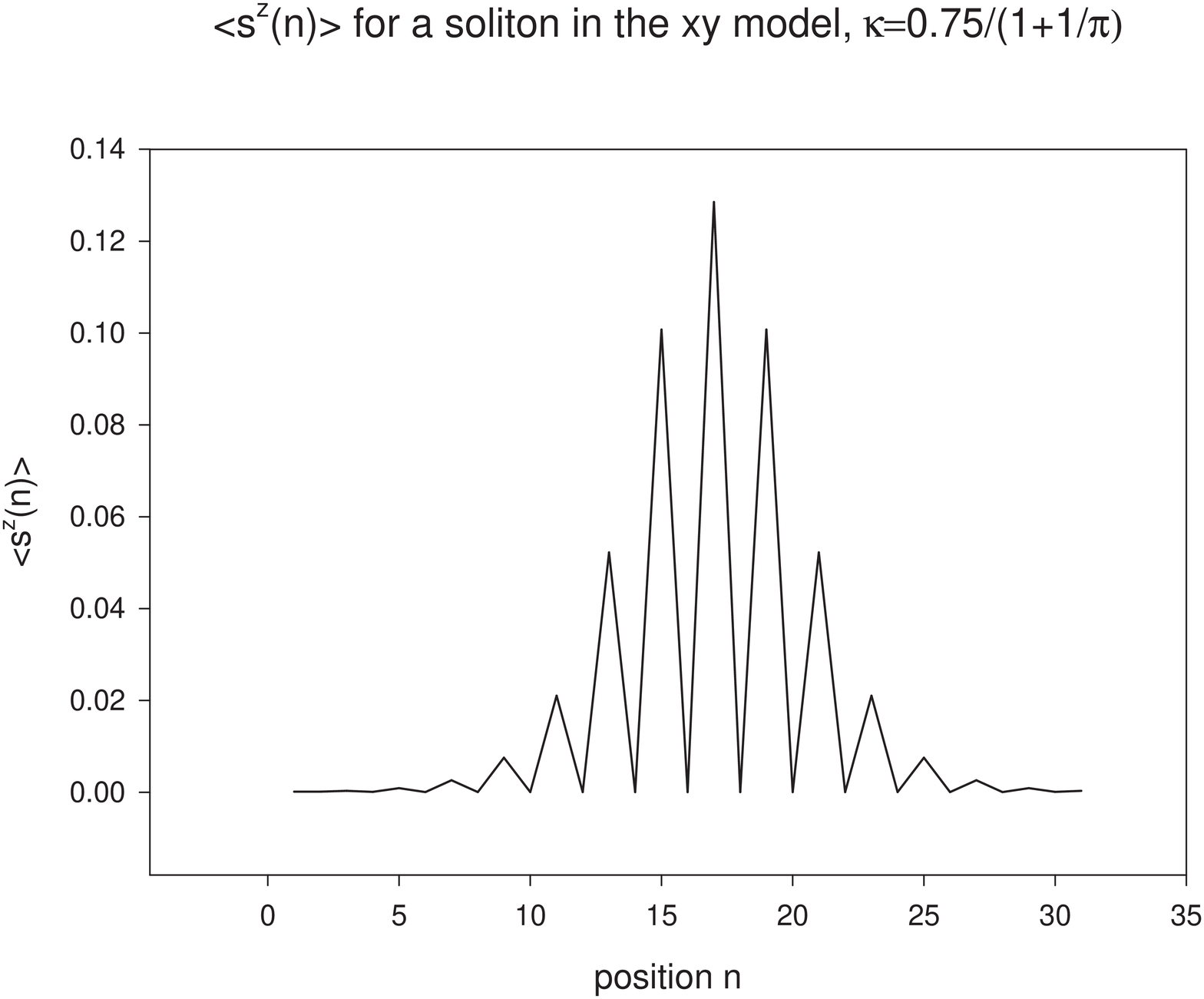,width=12cm}}\\
\caption{spin profile of a soliton in the xy Spin Peierls model}
\end{figure}

The magnetic profile of a soliton in the full Heisenberg spin Peierls 
model as obtained by DMRG and HF are given in fig.3. Clearly, this 
profile is quite different from that obtained in the xy Spin Peierls
model.

\begin{figure}
\mbox{\epsfig{file=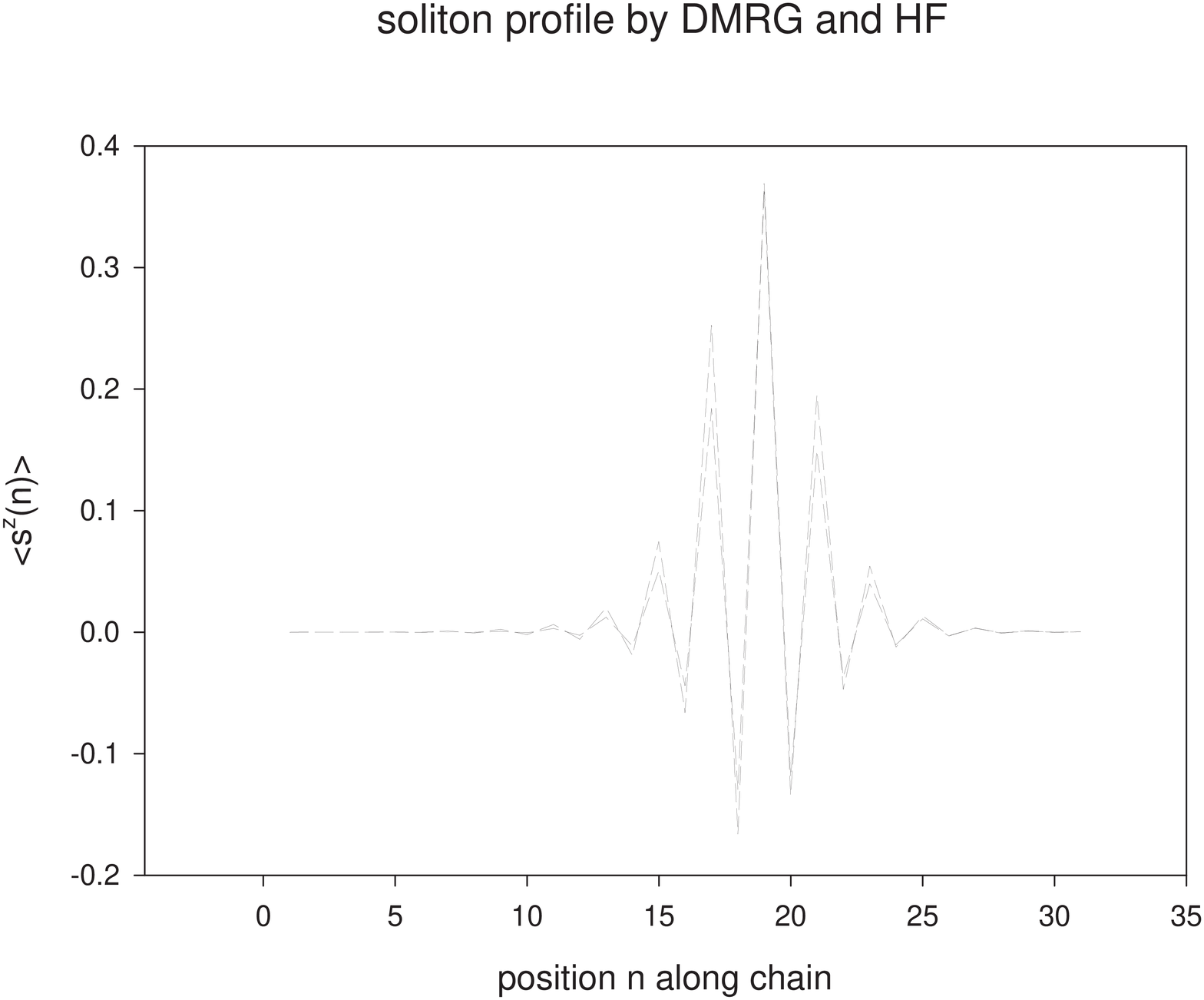,width=10cm}}\\
\caption{spin profile of a soliton in the Heisenberg Spin Peierls model
         according to HF and DMRG for $\kappa=0.75$}
\end{figure}

The DMRG result gives the symmetric profile in fig.3 and it is hard 
to differentiate between the two profiles.

\section{Solitons in the incommensurate phase and the NMR signal}

Since DMRG and HF give comparable results for the spin density
in the dimerised phase and since DMRG is difficult for
long inhomogenous chains, we shall use HF to estimate the distribution of
spins in the incommensurate phase\cite{BuzdinYann}.

\begin{figure}
\mbox{\epsfig{file=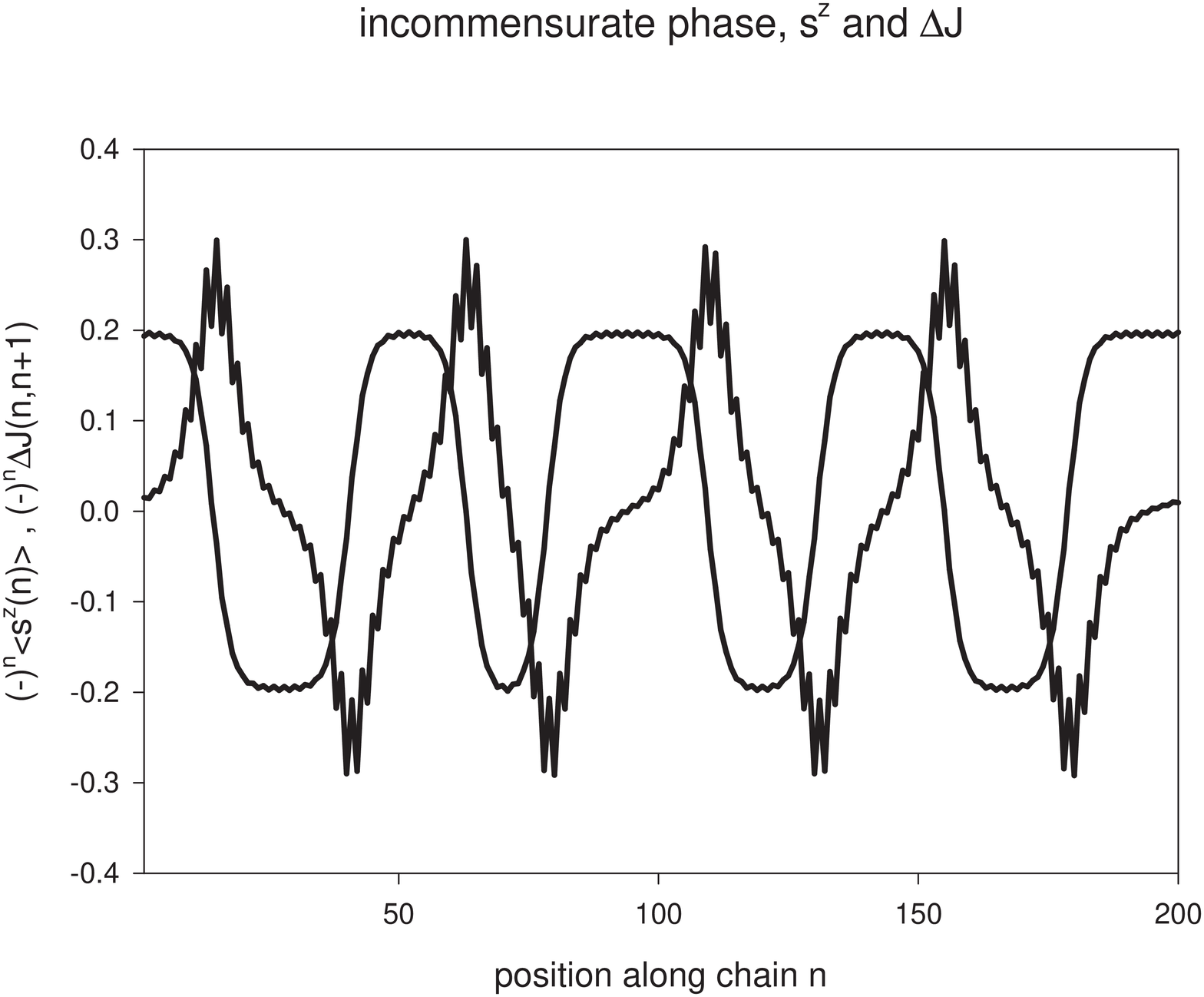,width=10cm}}\\
\caption{spins and lattice deformation in the incommensure phase for
         a chain of $L=200$, $\frac{\mu B}{J}=0.1$, $\kappa=1.5$}
\end{figure}

We crudely fix
the parameter $ \kappa $ by requiring that the period of the 
soliton structure be
comparable to that observed in the NMR experiments. Consider the
configuration of $\Delta J$ and $s^{z}$, multiplied by an oscillating factor,
in the incommensurate phase for $L=200$, 
$\kappa =1.5$, $\frac{\mu B}{J} = 0.1$ in fig.4.

\begin{figure}[h]
\mbox{\epsfig{file=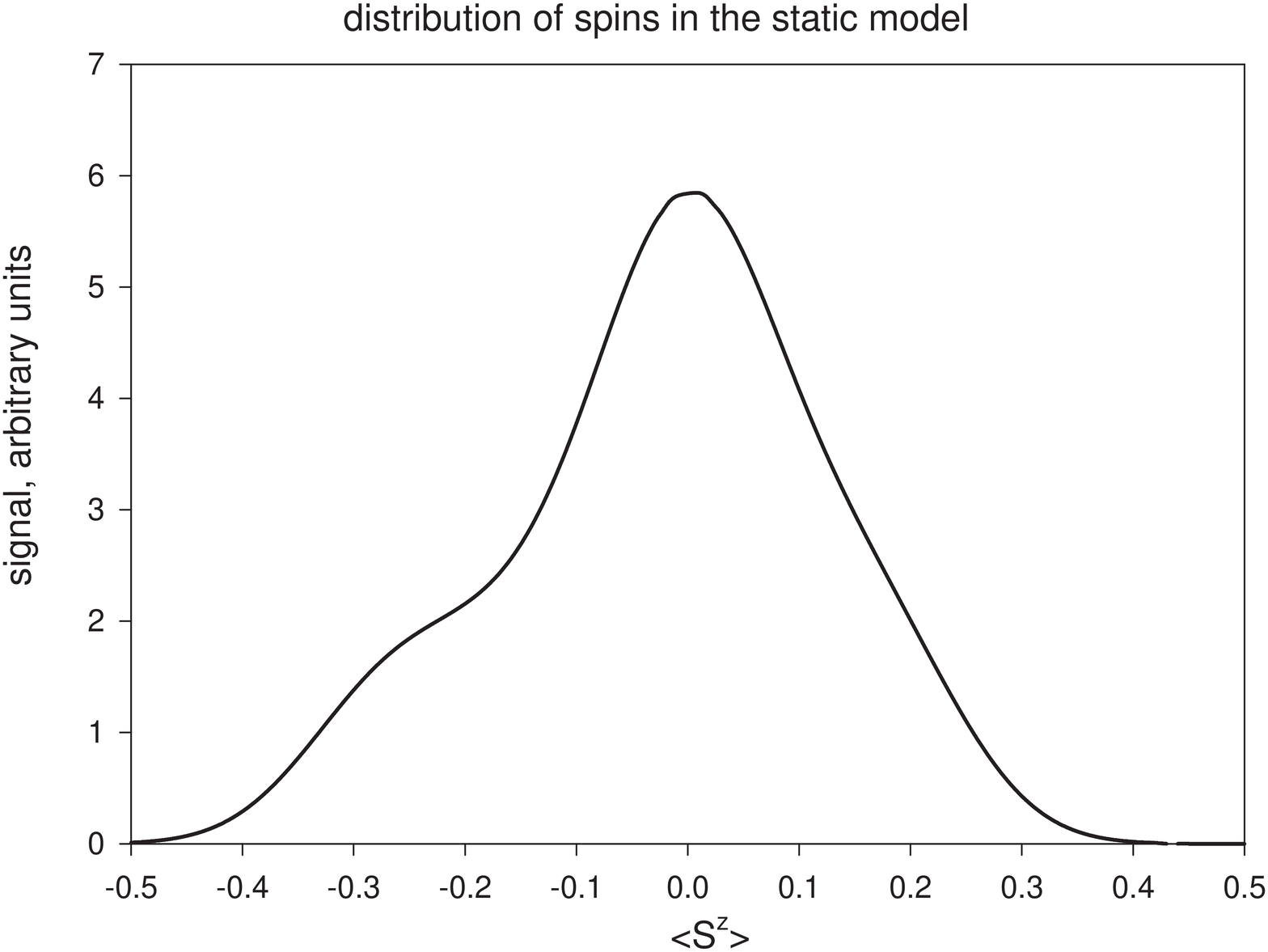,width=10cm}}\\
\caption{NMR signal from the static Spin Peierls model}
\end{figure}

The NMR data reflects the distribution of spins along the chain
or the number of points that carry a prescribed value of spin.
The configuration of spins as displayed in fig.4 (modulo an oscillating
factor), gives rise to the distribution of fig.5.
We see immediately that this distribution is incompatible with the NMR data of 
\cite{Berthier}: it contains positive and negative spins and the value of 
$s_{\max }^{z}-s_{\min }^{z}$ disagrees by a factor of $\sim 10$ with the
value quoted by \cite{Berthier}. 

So we are forced to conclude that the static Heisenberg spin Peierls model
for CuGeO is incompatible with the NMR data. The most plausible resolution 
of this

\begin{figure}[h]
\mbox{\epsfig{file=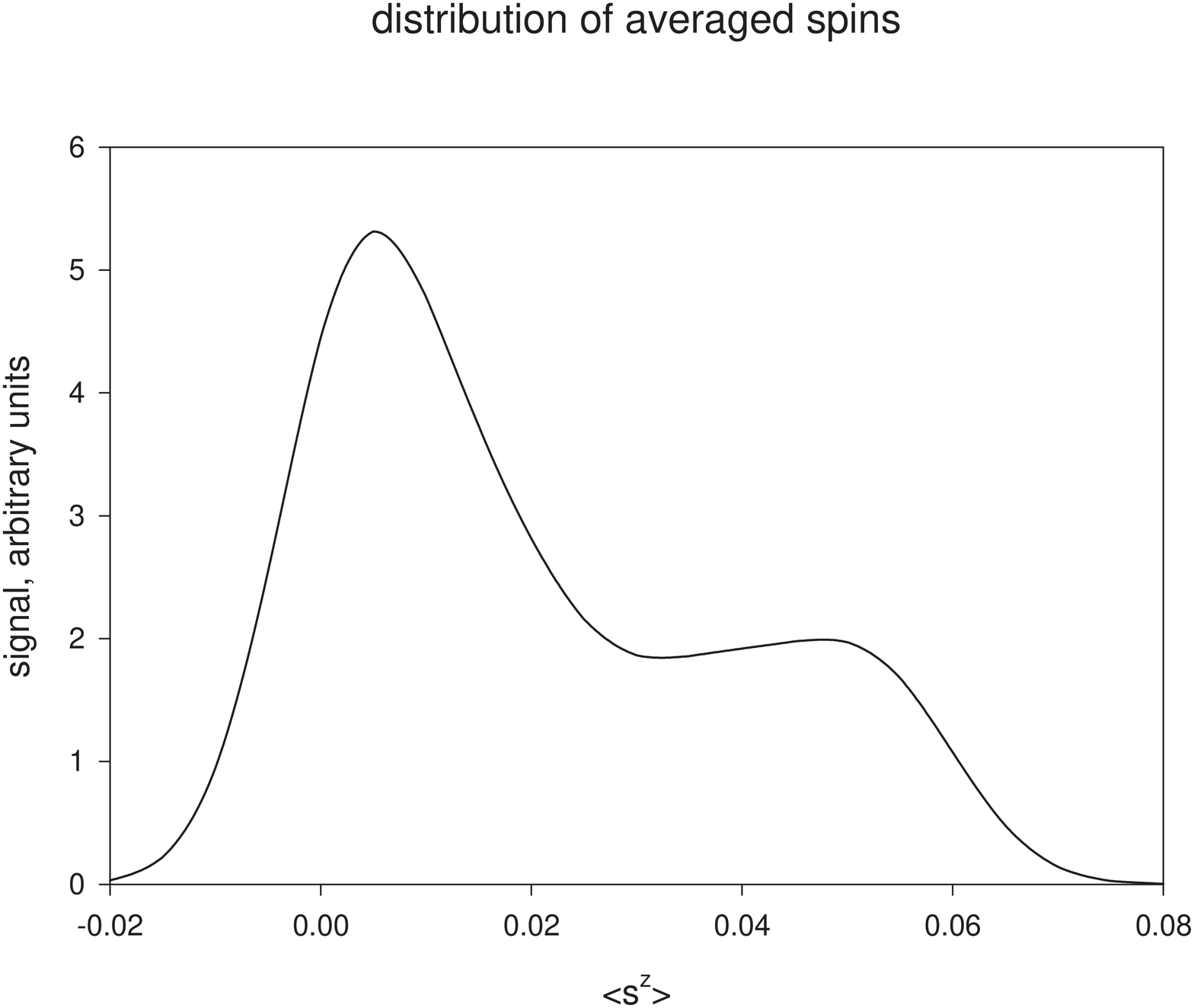,width=10cm}}\\
\caption{NMR signal from the distribution of averaged spins}
\end{figure}

discrepancy is that the motion of a nuclear spin in the NMR experiment
averages over the dynamics of the spin plus lattice system. To mimic 
such an average that
is, strictly speaking, outside of the domain of validity of our calculation,
we take averages $\frac{1}{2}\left( s_{n}^{z}+s_{n+1}^{z}\right) $
between
even and odd spins. The distribution of these average spins
that is given in fig.6 is qualitatively similar to the distribution of
spins in
the xy model, in that the averages have one sign only. 
The value of $s_{\max }^{z}-s_{\min }^{z}$ for the distribution
of averaged spins 
is of the same order as in the NMR data of \cite{Berthier},
but the agreement with the data is only on a crude and qualitative level.

\section{Conclusions}

We have calculated the magnetic structure of solitons in a static 
Heisenberg Spin Peierls model via DMRG and the Hartree Fock approximation.
Thanks to the fact that the renormalised xy model is not the HF 
approximation of the Heisenbergmmodel we found reasonable
agreement between HF and DMRG results in the dimerised phase. 

Within our static model, we obtained a spin distribution
in the incommensurate phase that is incompatible with the NMR data on 
CuGeO. We then assumed that the NMR data reflect averages in time over 
fluctuating spins and lattice deformations and mimicked such internal 
dynamics (that is beyond the scope of our static approach) by 
taking averages over even and odd spins. The resulting spin distribution
agrees qualitatively with the NMR data.

We conclude that the dynamics of the spin plus lattice
system must be taken into account in a proper theoretical description 
of the Spin Peierls transition of the CuGeO system.

\vspace{1cm}

{\bf Acknowledgements}: We are indebted to Alexander Buzdin for introducing
us to the Spin Peierls problem and for showing us the NMR results. We thank
Mladen Horvatic and Claude Berthier for comments on their data and Sergei
Brasovskii, Lev Bulaevski and Kazumi Maki for comments on the theory. We
also thank Goetz Uhrig for calling our attention to the QMC calculatios of
ref\cite{Feiguin} that we had been unaware of.

\section{\bf Appendix}

To correct an error in the litterature, we give the details of the Hartree 
Fock calculation in the dimerised phase
i.e. for prescribed and uniform $\Delta J$ with $B=T=0=<s_{n}^{z}>=\rho
_{n,n}-\frac{1}{2}=0$, see eq(\ref{HartreeFock}): 
\begin{eqnarray}
H_{eff} &=&\sum \frac{1}{2}K_{n,n+1}\psi _{n+1}^{+}\psi _{n}+h.c.  \label{HF}
\\
K_{n,n+1} &=&J_{n,n+1}*s_{n,n+1}\mbox{, }s_{n,n+1}=1-2\rho _{n,n+1}=1-2<\psi
_{n}^{+}\psi _{n+1}>  \nonumber \\
E(\rho ) &=&-\frac{1}{4}\sum J_{n,n+1}\left( s_{n,n+1}s_{n+1,n}-1\right) 
\nonumber
\end{eqnarray}
We are interested in alternating couplings: 
\begin{eqnarray}
J_{n,n+1} &=&1+\Delta J(-)^{n}  \label{alternation} \\
K_{n,n+1} &=&J_{n,n+1}s_{n,n+1}=K+\Delta K(-)^{n}  \nonumber \\
\Delta K &=&s\Delta J+\Delta s\mbox{, }K=s+\Delta J\Delta s  \nonumber \\
E(\rho ) &=&-\frac{1}{4}volume*\left( s^{2}+\Delta s^{2}+2\Delta Js\Delta
s-1\right)  \nonumber
\end{eqnarray}
Fourier transforming the effective Hamiltonian and diagonalising it
\begin{eqnarray}
H_{eff} &=&\frac{1}{2}\sum \left( K+\Delta K(-)^{n}\right) \psi
_{n+1}^{+}\psi _{n}+h.c..  \label{Diagonalising} \\
&=&\sum_{|p|<\pi /2}\Lambda (p)\left[ -\alpha ^{+}(p)\alpha (p)+\beta
^{+}(p)\beta (p)\right]  \nonumber \\
\Lambda &=&\sqrt{K^{2}\cos ^{2}p+\Delta K^{2}\sin ^{2}p}  \nonumber \\
\psi (p) &=&u(p)\alpha (p)+v^{*}(p)\beta (p)  \nonumber \\
\psi (p+\pi ) &=&v(p)\alpha (p)-u^{*}(p)\beta (p)  \nonumber
\end{eqnarray}
we find two consistency conditions: 
\begin{eqnarray}
\rho _{2n,2n+1} &=&<\psi _{2n}^{+}\psi _{2n+1}>=t+\Delta t
\label{consistency} \\
\rho _{2n+1,2n+2} &=&<\psi _{2n+1}^{+}\psi _{2n+2}>=t-\Delta t  \nonumber \\
t &=&<\sum_{|p|<\pi /2}e^{ip}\left[ u^{*}(p)u(p)-v^{*}(p)v(p)\right] =-\frac{%
1}{\pi }\left( {\bf K}-{\bf D}\right)  \nonumber \\
\Delta t &=&<\sum_{|p|<\pi /2}\left[ u^{*}(p)v(p)-v^{*}(p)u(p)\right]
e^{ip}=-\frac{\Delta K}{\pi K}{\bf D}  \nonumber \\
\kappa &=&\sqrt{1-\left( \frac{\Delta K}{K}\right) ^{2}}  \nonumber
\end{eqnarray}
At $T=B=0$ both conditions can be combined into one fixed point equation for 
$\frac{\Delta s}{s}$:

\begin{eqnarray}
s &=&1-2t\text{, }\Delta s=-2\Delta t \\
\Delta K/K &=&\frac{\Delta J+\Delta s/s}{1+\Delta J\Delta s/s}  \nonumber \\
\Delta s &=&\frac{2\Delta K}{\pi K}{\cal D}\mbox{, }s=1+\frac{2}{\pi }%
\left( {\cal K}-{\cal D}\right)  \nonumber \\
&\rightarrow &\frac{\Delta s}{s}=\frac{\frac{2\Delta K}{\pi K}{\cal D}}{1+%
\frac{2}{\pi }\left( {\cal K}-{\cal D}\right) }  \nonumber
\end{eqnarray}
From $\{\Delta s,s\}$ we can compute the ground state energy $E(\Delta J)$.
The astonishing quality of the HF calculation raises the challenge of going
beyond this approximation in a controlled way.


\begin{references}
\bibitem{BuzdinYann}  A systematic study of the incommensurate phase by DMRG
methods will be given by Yann Meurdesoif in collaboration with Alexander
Buzdin.
\end{references}
\end{document}